\documentstyle[12pt,a4wide,amsmath,amssymb,epsf]{article}

\newcommand{\be}{\begin{equation}}
\newcommand{\ee}{\end{equation}}
\newcommand{\bea}{\begin{eqnarray}}
\newcommand{\eea}{\end{eqnarray}}

\newcommand{\hf}{\frac{1}{2}}
\newcommand{\nn}{\nonumber\\}

\begin{document}

\begin{center}
{\large {\bf String cosmology from a novel renormalization perspective: 
perturbing the fixed point}}

\vspace{1cm}

{\bf Jean Alexandre} and {\bf Nikolaos E. Mavromatos} 

\vspace{0.3cm}

Department of Physics, King's College London, London  WC2R 2LS, England 
                                                                                                                             
\vspace{1cm}
                                                                                                                             
{\bf Abstract}
                                                                                                                             
\end{center}

We study small perturbations around the exactly marginal time-dependent string configuration of
[1], and demonstrate the lack of the appropriate linearization. This implies that 
this  configuration is an isolated fixed point of the $\alpha^{'}$ flow in the 
pertinent space of theories.

\vspace{2cm}

In \cite{AEM1}, the authors were interested in a time-dependent configuration of the bosonic string,
relevant to the description of a spatially-flat Robertson Walker Universe, with metric
$ds^2=-dt^2+a^2(t)(d\vec x)^2$, where $t$ is the time in the Einstein frame, and $a(t)$ is the scale factor.
It was shown that the following time-dependent configuration
\be\label{marginal}
S=\frac{1}{4\pi\alpha^{'}}\int d^2\xi\sqrt{\gamma}\left\{\gamma^{ab}\frac{\kappa_0}{(X^0)^2}\eta_{\mu\nu}
\partial_a X^\mu\partial_b X^\nu+\alpha^{'}R^{(2)}\phi_0\ln(X^0)\right\},
\ee
where $\kappa_0$ and $\phi_0$ are constants, does not get renormalized after quantization. It was then
conjectured that it satisfies Weyl invariance conditions in a non-perturbative way,
for any target space dimension $D$.
The corresponding scale factor was then shown to be a power law
\be
a(t)\propto t^{1+\frac{D-2}{2\phi_0}},
\ee
such that, if the following relation holds
\be\label{Minkowski}
D=2-2\phi_0,
\ee
the target space is static and flat (Minkowski Universe).

\vspace{0.5cm}

Our aim in this note is to check on the stability properties of the configuration (\ref{marginal})
under small perturbations (linearization, to be defined below).
Specifically, we shall consider small perturbations around the configuration (\ref{marginal}), with metric 
$g_{\mu\nu}$ and dilaton $\phi$ such that
\bea\label{newconfig}
g_{\mu\nu}(X^0)&=&\kappa(X^0)\eta_{\mu\nu}=\frac{\kappa_0}{(X^0)^2}\eta_{\mu\nu}\left(1+\xi(X^0)\right)\nn
\phi^{'}(X^0)&=&\frac{\phi_0}{X^0}\left(1+\varepsilon(X^0)\right),
\eea
where a prime denotes a derivative with respect to $X^0$
and $\xi<<1$, $\varepsilon<<1$. For this configuration, the beta functions, whose 
vanishing would guarantee world sheet Weyl invariance, are
to the lowest order in $\alpha^{'}$ \cite{AEM1},
\bea\label{beta}
\beta^g_{00}&=&-(D-1)\left(\frac{\kappa^{'}}{2\kappa}\right)^{'}+2\phi^{''}
-\frac{\kappa^{'}}{\kappa}\phi^{'}+{\cal O}(\alpha^{'})\nn
\beta^g_{ij}&=&\delta_{ij}\left\{\left(\frac{\kappa^{'}}{2\kappa}\right)^{'}
+(D-2)\left(\frac{\kappa^{'}}{2\kappa}\right)^2-\frac{\kappa^{'}}{\kappa}\phi^{'}\right\}+{\cal O}(\alpha^{'})\nn
\beta^\phi&=&\frac{D-26}{6\alpha^{'}}-\frac{\phi^{''}}{2\kappa}-\frac{(D-2)}{4}\frac{\kappa^{'}}{\kappa^2}\phi^{'}
+\frac{(\phi^{'})^2}{\kappa}+{\cal O}(\alpha^{'}).
\eea
Substituting the perturbation (\ref{newconfig}) and keeping only the linear terms in $\xi$ and $\varepsilon$, leads to
\bea
\beta^g_{00}&=&-\frac{D-1}{(X^0)^2}+\frac{1}{(X^0)^2}\left\{
-\frac{D-1}{2}(X^0)^2\xi^{''}+\phi_0X^0(2\varepsilon^{'}-\xi^{'})\right\}+{\cal O}(\alpha^{'})\\
\beta^g_{ij}&=&\delta_{ij}\frac{D-1+2\phi_0}{(X^0)^2}+\frac{\delta_{ij}}{(X^0)^2}\left\{
\frac{(X^0)^2}{2}\xi^{''}-(D-2)X^0\xi^{'}+\phi_0(2\varepsilon-X^0\xi^{'})\right\}+{\cal O}(\alpha^{'})\nn
\beta^\phi&=&\frac{D-26}{6\alpha^{'}}+\frac{\phi_0}{2\kappa_0}(D-1+2\phi_0)\nn
&&+\frac{\phi_0}{2\kappa_0}\left\{
-X^0\left(\varepsilon^{'}+\frac{D-2}{2}\xi^{'}\right)+(D-1)(\varepsilon-\xi)
+2\phi_0(2\varepsilon-\xi)\right\}+{\cal O}(\alpha^{'}).\nonumber
\eea
We are looking for solutions $\xi,\varepsilon$ which, at this order in $\alpha^{'}$,
do not change the homogeneity in $X^0$ of the beta functions, and therefore satisfy
\bea\label{equadiffs}
0&=&-\frac{D-1}{2}(X^0)^2\xi^{''}+\phi_0X^0(2\varepsilon^{'}-\xi^{'})\nn
0&=&\frac{(X^0)^2}{2}\xi^{''}-(D-2+\phi_0)X^0\xi^{'}+2\phi_0\varepsilon\nn
0&=&-X^0\left(\varepsilon^{'}+\frac{D-2}{2}\xi^{'}\right)+(D-1+4\phi_0)\varepsilon
-(D-1+2\phi_0)\xi
\eea
In what follows, we are interested in a Minkowski target space, and consider therefore the 
situation where $D=2-2\phi_0$, such that
\bea\label{equadiffsbis}
0&=&\left(\phi_0-\hf\right)(X^0)^2\xi^{''}+\phi_0X^0(2\varepsilon^{'}-\xi^{'})\nn
0&=&\frac{(X^0)^2}{2}\xi^{''}+\phi_0X^0\xi^{'}+2\phi_0\varepsilon\nn
0&=&-X^0\left(\varepsilon^{'}-\phi_0\xi^{'}\right)+(1+2\phi_0)\varepsilon-\xi
\eea
The structure of these equations shows that the only possible solution is a power law for $\varepsilon$ and
$\xi$. We then assume the following $X^0$-dependence, consistent with our linearization procedure,
\be
\varepsilon=\frac{A}{(X^0)^a}~~~~~~~~\xi=\frac{B}{(X^0)^a},
\ee
where $A,B,a$ are constants to be determined. Using the first two of the equations (\ref{equadiffsbis}), 
one arrives at
\be
0=4\phi_0+a^2-1.
\ee
And then, using the last two of equations (\ref{equadiffsbis}), one obtains 
\be
1+a^2=0,
\ee
which has no real solution.

We conclude that no linearization is possible around the Minkowski Universe described 
by the exactly marginal configuration (\ref{marginal}), thereby suggesting that the 
latter is an isolated fixed point of the $\alpha^{'}$ flow in the pertinent space of quantum theories.

\end{document}